\documentclass[prb,twocolumn,floatfix,showpacs,superscriptaddress]{revtex4}
\usepackage{graphics}
\usepackage{amsmath}
\usepackage{epsfig}
\usepackage{color}
\usepackage{times}

\begin{document}
\preprint{3HEP/123-qed}
\title{Boundary effects in the density-matrix renormalization group calculation}
\author{Naokazu Shibata}
\affiliation{Department of Physics, Tohoku University, Sendai 980-8578, Japan}
\author{Chisa Hotta}
\affiliation{Department of Physics, Kyoto Sangyo University, 
Kyoto 603-8555, Japan}
\date{\today}
\begin{abstract}
We investigate {\it the boundary effect} of the density matrix 
renormalization group calculation (DMRG), which is an artifactual 
induction of symmetry-breaking pseudo-long-range order 
and takes place when the long range quantum fluctuation cannot 
be properly included 
in the variational wave function due to numerical limitation. 
The open boundary condition often used in DMRG suffers from the 
boundary effect the most severely, which is directly reflected 
in the distinct spatial modulations of the local physical quantity. 
By contrast, the other boundary conditions such as the periodic one or 
the $\sin^2$-deformed interaction [A. Gendiar, R. Krcmar, and T. Nishino, 
Prog. Theor. Phys. {\bf 122},  953 (2009)] keep spatial homogeneity, 
and are relatively free from the boundary effect.
By comparing the numerical results of those various boundary conditions, 
we show that the open boundary condition sometimes gives unreliable 
results even after the finite-size scaling. We conclude that the 
examination of the boundary condition dependence is required besides 
the usual treatment based on the system size or accuracy dependence 
in cases where the long-range quantum fluctuation is important.
\end{abstract}
\pacs{71.10.Hf, 71.27.+a, 71.10.-w}
\maketitle
\narrowtext 

\section{Introduction}
The density matrix renormalization group (DMRG) proposed by White\cite{white92} 
is by now one of the most powerful numerical tools to determine the ground 
states and low-energy excitations\cite{review1,review2} of the correlated 
quantum many-body systems in low dimensions. 
Accumulated studies revealed that the DMRG has advantages when the system 
is not critical (gapped), where one could effectively reduce the 
number of bases states
without sacrificing the numerical accuracy. 
The open boundary condition (OBC) usually used in the DMRG analysis 
also works to 
reduce the number of basis,  
which was pointed out by White in the first stage.\cite{white92} 
In fact, compared to the periodic boundary condition (PBC), OBC has a smaller 
number of joint points between the blocks (subsystems) 
than PBC (one and two points, respectively), and thus has smaller 
entanglement entropy which follows a concept of 
area law:\cite{area-law} the smaller the size of the boundary between the 
two blocks, the smaller the entanglement entropy one finds between them. 
However, one shall have to be careful in this context because the 
existence of the open edge points does not only indicate the reduced 
entanglement, 
but means the breaking of the translational symmetry which leads to 
two artifacts in critical (gapless) systems.
One is to favor a particular fraction of ground-state manifolds which 
are equivalently degenerate in the bulk limit. 
The other is {\it to mix partially the states which have different 
symmetric properties from the ground-state manifold}. 
The former effect does not modify the ground-state nature (only chooses 
a pure state) and is equivalent to introducing the infinitesimal 
symmetry-breaking field in large systems (see Sec.~II A for details). 
However, the latter may sometimes allow some low-energy excited state 
to overwhelm the original ground state. 
For example, when the correlation length is longer than the typical 
system size, 
the oscillation induced by the open boundary 
may open an apparent gap, particularly when the number of states kept is 
not large enough. This is what we call the ``boundary effect.'' 
\par
The portion of boundary is order 1 while the system is order $L$, 
and by the finite-size scaling, the boundary effect
is expected to decay faster than $1/L$ on an average. 
Therefore, by the proper finite-size scaling, one could in principle get 
rid of the boundary effect, and the correlation with the largest 
length scale is considered to be the one characterizing the ground state. 
In fact, one can take advantage of the OBC in order to examine the nature 
of orders or correlations: Due to the boundary edges, the wave function 
loses the translational symmetry and the two-point correlation of the 
largest length scale explicitly appears as a spatial modulation of 
{\it local physical quantities}. 
This treatment was actually applied to DMRG by the present authors.
In the one-dimensional Kondo lattice model (KLM) 
the two open edges are regarded as impurities which induce the Friedel 
oscillations, 
and by measuring the structural factors of the charge 
or spin amplitudes, the authors detected the wave numbers, $2k_F$ or $4k_F$, 
of the Tomonaga-Luttinger liquid state.\cite{shibata96}
Also in two-dimension, the competition of several different types of 
long range orders 
with different spatial periodicity is detected by the spatial modulation of 
particle density in the fermionic model on an anisotropic triangular 
lattice under the chemical 
potential on system edges.\cite{nishimoto09}
Such analysis is, however, applicable only when the basic nature of the 
ground state is already understood; namely, it {\it clarifies the 
details of the already known (stable) ground state but not the ground 
state itself}. 
Moreover, the analysis requires a set of results with high enough 
accuracy as well as a parameter region where the scaling law is safely adopted. 
Once either of the above two is not fulfilled, one may no longer obtain 
a reliable result, which is often the case in DMRG, since there is an 
upper bound of the number of states kept, in practice. 
\par
The present paper discusses how to classify systematically the state 
which does or does not break any symmetry of the original 
Hamiltonian within a usual numerical accuracy of DMRG. 
We examine how the critical behavior of quantum many-body states at 
finite system size is influenced by the boundary condition by choosing 
two characteristic examples: a well-known symmetry-broken long-range
 order in a spin chain, and strongly correlated electronic states with 
extremely long correlation length. 
We consider the modified or deformed OBC, as well as PBC, and demonstrate 
that the expectation value of the local operator depends severely on 
boundary conditions so that they cannot be a reliable measure. 
We finally see that in order to conclude the presence of long-range order, 
it is necessary to confirm that {\it the two-point correlations remain finite 
toward the bulk limit irrespective of the boundary conditions}. 
\section{Boundary conditions}

\subsection{Symmetry-breaking long-range orders}
\label{sec2:tasaki}
We consider a class of long-range orders due to translational symmetry 
breaking; i.e., the order operators do not commute with the quantum 
Hamiltonian of crystals which usually keeps the translational symmetry. 
It is known that even when such spontaneous symmetry breaking occurs in 
the bulk limit, quantum fluctuation ``obscures'' the breaking of symmetry 
at finite system size and one finds a unique ground state with perfect 
symmetry.\cite{tasaki94,horsch88}
\par
Let us briefly follow the contexts of Ref.~\onlinecite{tasaki94} by Koma 
and Tasaki, which is based on two different kinds of measures, 
\begin{eqnarray}
m_s^{\rm } & = & \lim_{B\rightarrow 0} \lim_{L\rightarrow\infty} m_s(B,L), 
\label{op1} \nonumber \\
&& m_s(B,L) = \frac{1}{L} \big\langle \sum_i  O_i (B,L) \big\rangle, \\
\sigma^{\rm } & = & \lim_{L\rightarrow\infty} \sigma(L), \nonumber \\
&& \sigma(L) = \frac{1}{L} \sqrt { \Big\langle 
\Big(\sum_i  O_i (B=0,L)\Big)^2 \Big\rangle }, 
\label{op2}
\end{eqnarray}
where $\langle O_i(B,L) \rangle$ is the expectation value (trace) of local 
operator $O_i(B,L)$ on the $i$th site 
under the symmetry-breaking field $B$ and for system size $L$. 
Hereafter, we use mere $O_i$ as the one with $B=0$ and finite $L$ 
except otherwise noted. In the Heisenberg antiferromagnet, $m_s^{\rm }$ 
is the staggered magnetization obtained by applying the infinitesimally 
small symmetry-breaking field, $B\rightarrow 0$, 
whereas the latter is the conventional long-range order parameter 
which consists of two-point correlation functions, 
$\langle O_i O_j \rangle$, between sites $i$ and $j$. 
Koma and Tasaki proved that 
\begin{equation}
m_s^{\rm } \ge r \sigma^{\rm }, 
\label{m_sa}
\end{equation}
holds for constant $r$ at all temperature for various types of models 
with long-range order [$r=1$ holds in general, but a higher symmetry 
gives stronger bounds, e.g., $r=\sqrt{3}$ for the SU(2) Heisenberg 
antiferromagnet\cite{tasaki94,tasaki93}]. 
Here, $\sigma > 0$ means that the two-point correlation function 
does not decay to zero in the bulk limit, 
$\langle O_i O_j \rangle \ne 0$ for $|i-j|\rightarrow \infty$. 
Therefore, Eq.(\ref{m_sa}) indicates that $\sigma > 0$ guarantees 
the existence of symmetry breaking represented by $m_s > 0$. 
\par
We now interpret Eq.(\ref{m_sa}) to the numerical analysis on a finite 
size system at zero temperature. 
Note that the following Eqs.~(\ref{mpbc})--(\ref{sobc}) and 
conditions (I)--(III) 
are conjectures, which are derived logically in the following part of 
this section. 
Since the finiteness of the system is characterized by the presence of 
boundaries, we first discuss the role of boundary conditions. 
When the system does not have any open edges, e.g., periodic (PBC) 
or antiperiodic (APBC) boundaries, the translational symmetry in
the Hamiltonian ``obscures'' the breaking of translational symmetry
at finite $L$, and we find $\langle O_i (L<\infty) \rangle_{\rm pbc}=0$. 
Thus we need to include the infinitesimal symmetry-breaking field $B$,
or analyze the two-point correlation function $\sigma(L)_{\rm pbc}$,
which coincides with $\sigma$ in the limit of large $L$:
$\displaystyle \lim_{L\rightarrow\infty} 
\sigma(L)_{\rm pbc} = \sigma$.
\par 
In the case of open boundary condition, 
the coupling between the 1st and the $N$th sites in PBC is missing
and the translational symmetry is broken in the Hamiltonian.
This works to discriminate the two edge bonds energetically from 
the rest of the bonds in the system, 
which approximately corresponds to placing the effective external 
field on the edge bonds.\cite{lou00} 
Such effective local field induces oscillation of local 
correlations, which is absent in PBC. 

We basically confine ourselves to the case where the period of the 
oscillation is compatible with the interval between edge bonds; 
e.g., if the induced oscillation has twofold periodicity, the number 
of bonds, $L-1$, must be odd. 
Then the open boundary condition corresponds to a locally introduced 
symmetry-breaking ``field,'' $B_{\rm ob}$, which is coupled to 
the local operator $O_i = o_{i+1}o_{i+2}-o_{i}o_{i+1}$ at the edge sites, 
$i=1$ and $N$, 
where $o_i$ is the operator acting on the $i$th site.
The effect of $B_{\rm ob}$ decays as it propagates toward the system 
center from both edges, 
which is reflected in the gradual decrease of the oscillation amplitude 
of $\langle O_i\rangle$ with $i\rightarrow N/2$. 
Then, we expect $m_s(L)_{\rm obc} \ge m_s$. 
The two-point correlation functions $\langle O_i O_j \rangle_{\rm obc}$ 
are those under the effective symmetry breaking ``field,'' 
so that we shall find $\sigma(L)_{\rm obc} \ge \sigma$. 
\par
Basically the portion of boundary is order 1 while the system is order $L$,
and taking the limit of $L\rightarrow \infty$ corresponds qualitatively 
to having $B_{\rm ob}\rightarrow 0$ as in Eq.(\ref{op1}).
Thus both of the above two boundary conditions (PBC and OBC) lead to 
the same conclusion, 
\begin{eqnarray}
m_s &=& \lim_{B\rightarrow 0}  \lim_{L\rightarrow\infty} m_s(B,L)_{\rm pbc} \label{mpbc} \\
&=& \lim_{L\rightarrow\infty} m_s(B=0,L)_{\rm obc}, \\
\sigma &=& \lim_{L\rightarrow\infty} \sigma(L)_{\rm pbc }  \\
 &=& \lim_{L\rightarrow\infty} \sigma(L)_{\rm obc}, \label{sobc}
\end{eqnarray}
with $m_s\ge \sigma$.
We therefore conclude that when $\sigma >0$, we always find $m_s > 0$. 
\par
However, the numerical approximation, e.g., restricting the 
number of basis to $m$ in DMRG, sacrifices the inclusion of long-range 
quantum fluctuation to some extent. 
The OBC does not only work as an ``effective symmetry breaking field,'' 
but its locality induces a nonuniformity of the quantum fluctuation and 
mixes the states which have different symmetry from the original pure state. 
While such additional effects may be weakened as $L\rightarrow\infty$, 
incomplete long-range quantum fluctuations sometimes cause harmful effect 
on the original ground-state symmetry as we will see shortly. 
Therefore, the practical procedure to systematically get rid of the 
artifact of the boundary condition is required. 

The above-mentioned nonuniformity under OBC is reflected in 
$\langle O_i\rangle$ and $\langle O_n O_{n+i}\rangle$, and their 
deviation from the one in the uniform system 
is enhanced near the open edges. 
Since we need to analyze the systematic behavior of these quantities in 
large systems, 
we examine them at each site $i$ and $n$ instead of taking their mean 
value over the whole system
as in Eqs.~(\ref{op1}) and (\ref{op2}). 
Then, those of site $i$ and $n$ away from the edges are adopted which 
should fulfill 
$\lim_{L\rightarrow\infty} \langle O_i\rangle \gtrsim r 
\lim_{L\rightarrow\infty} \langle O_n O_{n+i}\rangle$ 
as in Eq.~(\ref{m_sa}). 
\par
With this in mind, we propose {\it the necessary conditions} the numerical 
results must fulfill in order to safely conclude the breaking of 
translational symmetry in the bulk limit:
\begin{description}
\item[\rm (I)] the two-point correlation function
$\langle O_n O_{n+i}\rangle$ remains finite for 
$i\rightarrow L/2$, after the $L$ and $m$ scaling, 
\item[\rm (II)] the conclusion obtained by (I) does not depend on the 
boundary conditions, including those with and without symmetry breaking, 
\item[\rm (III)] as $\langle O_i\rangle$ of finite systems becomes uniform 
under the variation of the boundary 
conditions, the corresponding two-point correlation function 
$\langle O_n O_{n+i}\rangle$ approaches  
the value of the pure state in the bulk limit. 
\end{description}
The rest of the paper is devoted to the numerical ``proof'' of this 
conjecture or proposal in the representative models in one dimension. 
\par
The boundary conditions we consider are classified into those which keep the 
translational symmetry and those which do not. 
As a typical boundary condition of the former class, 
we deal with the system with ``deformed interactions'' 
recently proposed by Gendiar, Krcmar, and Nishino,\cite{nishino09} 
as well as the PBC and APBC. 
The representative condition of the latter class is the OBC. 
Besides, we consider the cases referred to as {\it modified open boundary}, 
which is the open boundary with potentials on edge sites or edge bonds 
first adopted by White and Huse.\cite{white93}
As we discussed in this section, the open boundary works as 
an effective field on edges, 
which brings the spacial nonuniformity of the quantum states. 
This artificial effect can be suppressed by adjusting the 
bond strength or potentials on 
edges by hand, and thus one can tune the degree of inhomogeneity 
by the ``modification of the open boundary.'' 
\par
We finally comment on the cases where the period of oscillation of local 
quantity $\langle O_i \rangle$ does not match the length of the 
system $L$. For example, in the case of dimer order, 
the local operator $O_i$ is defined as 
$ O_i = {\bf S}_{i+1}\cdot{\bf S}_{i+2} -{\bf S}_{i}\cdot{\bf S}_{i+1}$, and 
the twofold periodic oscillation appears in the 
nearest-neighbor spin-spin correlation. 
In the odd-$L$ system with PBC, a kink, namely a twisting 
oscillation emerges in the system. 
Since the PBC keeps translational symmetry, the wave function should be 
the superposition of $L$ different wave functions, each with a kink on 
the $i$th ($i=1\sim L$) bond. 
Then, $\langle O_i \rangle$ is uniform but is suppressed from the 
one without a kink by the order of $1/L$, which recovers the bulk 
value when $L\rightarrow\infty$. 

When the system of odd $L$ has an open boundary, this kink is confined 
to the center site as $\langle O_{L/2} \rangle \sim 0$. 
This is because the dimer bond is pinned the strongest at both edge bonds, 
and the oscillations of $\langle O_i \rangle$ starting from these strong 
edge bonds interfere at the center and form a kink.\cite{sp} 
The effect of this kink on $\langle O_{L/2} \rangle$ cannot be excluded 
by the $m$ or $L$ scaling, since it is locked at the center of the system. 
The artifact of kink on the value of $\langle O_n O_{n+i}\rangle$ 
is also nontrivial, and less easy to get rid of compared to the 
usual boundary effect of OBC with even $L$.  
We show in Sec.~IV that $\langle O_n O_{n+i}\rangle$ in the dimer state
with a kink in OBC finally approaches the value without a kink with 
phase shift $\pi$, 
which means that (II) basically holds regardless of the presence
or absence of a kink. 
In this sense we can judge the presence of the dimer order even in the 
system of odd $L$ with a kink.\cite{kink} 

In the above semiclassical picture, the kink is localized on a single bond. 
However, in the quantum systems we deal with, the kink spreads over a 
certain length scale due to quantum fluctuation, and 
this length scale depends on the stiffness of the dimer order. 
Still, if we take $L$ enough larger than this lengthscale, the same 
discussion holds.

\subsection{Modified open boundary}
\label{sec2:modify}
The analysis with ``modified open boundary'' in DMRG was first adopted 
in the $S=1$ Haldane chain with $S=1/2$ spins on the edge sites by 
White and Huse.\cite{white93} 
To evaluate the Haldane gap of infinite systems, 
they adjusted the coupling, $J_{\rm end}$, between the $S=1/2$ edge 
spin and the 
neighboring $S=1$ spin, and minimized the nonuniformity of the ground state 
away from the edge. 
\par
In the present paper, we start from the usual OBC and modify the 
amplitude of bond interactions on both ends or place the 
potential on both edge sites to analyze the boundary condition
dependence of the ground state. 
As for the spin system, the bonds, $J_{\rm edge}$, 
on both left and right edges are varied.
If we place the strong antiferromagnetic bond on one edge,
the singlet correlation on that edge is enhanced, 
and as a result, the correlation on the neighboring bond
decreases, which enhances or reduces the inhomogeneity of the ground state. 
In a similar manner, the electronic states can be tuned by 
modifying the hopping energy of edge bonds, $t_{\rm edge}$, 
or by placing the chemical potential on edge sites, $\mu_{\rm edge}$. 
One can minimize the nonuniformity of the electron density
by adjusting the value of $\mu_{\rm edge}$. 
Furthermore if one uses different values of $\mu_{\rm edge}$ between the two edges, 
the conduction electrons shift to left or right, 
and the position of the center of mass of electrons can be controlled.
\par
Besides the system with even number of $L$, we deal with odd number of 
$L$ to shift the center of mass of electrons smoothly along the 1D chain in OBC and at the same time 
keeping electron density constant away from the boundary. 
Although the average electron filling factor $\rho=N_e/L$ ($N_e$ denotes the 
electron number) slightly deviates from the one we have in the even-$L$ case, 
we can pin the excess electrons or holes to the boundary by
tuning the value of $\mu_{\rm edge}$. 

\begin{figure}[tbp]
\begin{center}
\includegraphics[width=7.5cm]{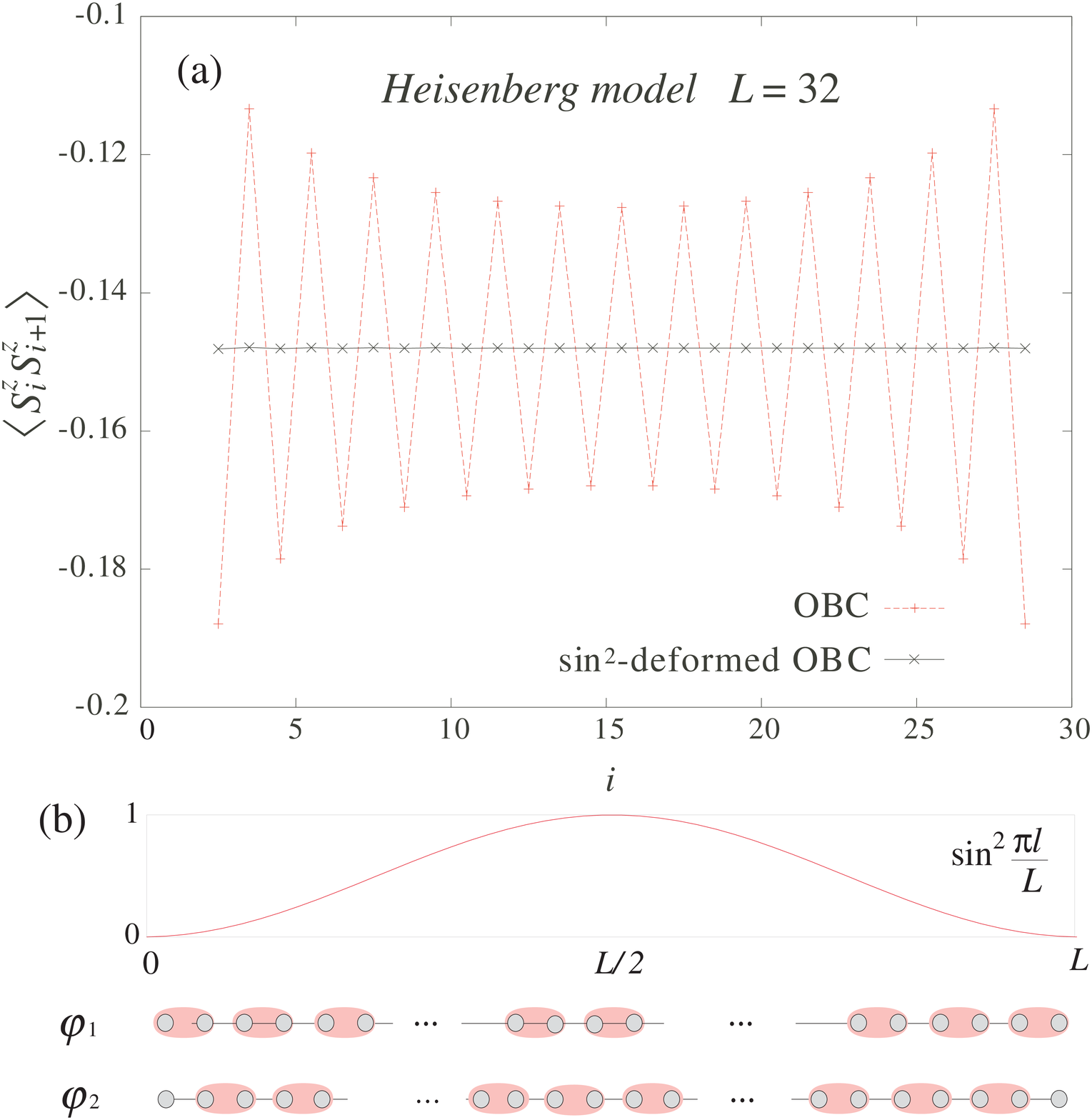}
\end{center}
\caption{
(a) Site dependence of the nearest-neighbor correlation, 
$\langle S_i^zS_{i+1}^z\rangle$, in the $S=1/2$ Heisenberg model with OBC 
and $\sin^2$ deformation at $L=32$ and $m=200$. 
(b) The functional form of the $\sin^2$ deformation in Eq.(\ref{deform}), 
and the two types of wave function of the dimer state, $\varphi_1$ and 
$\varphi_2$, 
which have the same energy. }
\label{f1}
\end{figure} 
\subsection{Deformed interactions}
\label{sec2:deform}
Recently, Gendiar, Krcmar, and Nishino proposed an unprecedented analysis to 
get rid of the boundary effect by deforming the interaction strength of the 
system following the $\sin^2$ function which decreases from the 
center toward both ends of the 
system.\cite{nishino09,ueda10,iharagi10,ueda09,gendiar11}
\par
In this paper we use the following deformed Hamiltonian:
\begin{eqnarray}
{\mathcal H} &=& \sum_{n=1}^{L-1} 
\sin^2\left(\frac{\pi n }{L}\right) g(n,n+1) \nonumber \\ 
& & +\; \sum_{n=1}^{L-2} 
\sin^2\left(\frac{\pi (n+1/2)}{L}\right) h(n,n+2) \nonumber \\ 
& & +\; \sum_{n=1}^{L} \sin^2\left(\frac{\pi (n-1/2)}{L}\right) u(n),
\label{deform}
\end{eqnarray}
where $g(n,n+1)$ is the nearest-neighbor interaction, 
 $h(n,n+2)$ is the next-nearest-neighbor interaction,
and $u(n)$ includes the on-site interaction and potential. 
The uniform chemical potential, $\mu$, is also included as $u(n)$ which 
is deformed by the prefactor and becomes nonuniform.\cite{gendiar11} 
Figure~\ref{f1} shows the site dependence of the nearest-neighbor 
spin-spin correlation of the $S=1/2$ Heisenberg model,
$g(n,n+1)=J\ {\bf S}_n \!\cdot \!{\bf S}_{n+1}$,
under the usual OBC and under the deformation of interaction. 
One finds that the usual OBC induces a twofold oscillation which 
decays slowly toward the system center. 
The oscillation is fully suppressed by the deformation and 
the local quantity becomes site independent. 
\par
Here, we give an interpretation on these ``site-independent'' results. 
Consider the system with even $L=2N$. 
Then, one finds the following relation: 
\begin{eqnarray}
 \sum_{l=1}^{N} J \sin^2\frac{\pi (2l) }{L}
=\sum_{l=1}^{N} J \sin^2\frac{\pi (2l-1) }{L}, 
\label{e_spehere}
\end{eqnarray}
which means that the sum of the coupling constant is the same 
between even bonds $(2l)$ and odd bonds $(2l-1)$ 
[although the total number of even bonds with finite coupling
is less than that of odd bonds because $\sin^2(2N\pi/L)=0$, 
which corresponds to the missing bond in OBC].
Suppose that we have translational symmetry broken dimer states 
with twofold periodicity, which are represented by two different types 
of wave functions, $\varphi_1$ and $\varphi_2$ as shown in the lower panel 
of Fig. \ref{f1}; 
$\varphi_1$ consists of dimers on even bonds and $\varphi_2$
consists of those on odd bonds.
If either of $\varphi_1$ and $\varphi_2$ has lower energy than the 
other, that state is selected as the ground state. 
However, due to the relation in Eq.(\ref{e_spehere}), $\varphi_1$ and 
$\varphi_2$ have the same energy,  
and thus the translationally symmetric wave function 
$(\varphi_1 \pm \varphi_2)/\sqrt{2}$ can be  
constructed, which has the same ground-state energy. 
Hikihara and Nishino recently calculated the electronic system with this 
$\sin^2$ deformation and found that 
the overlap with the PBC wave function is almost 1.\cite{hikihara11} 
This result was in fact supported analytically for the noninteracting 
case.\cite{katsura11} 
In the results of free fermionic model,\cite{nishino09} the Heisenberg chain 
[Fig.~\ref{f1}(c)], 
and $J_1$-$J_2$ models, which we will see shortly, 
the translational symmetry is 
indeed recovered by the deformation. 
Therefore, we consider that the $\sin^2$ deformation  
is a condition to have the translationally 
symmetric ground state at least for simple single-band short-range 
interacting systems. 
\begin{figure}[tbp]
\begin{center}
\includegraphics[width=7.5cm]{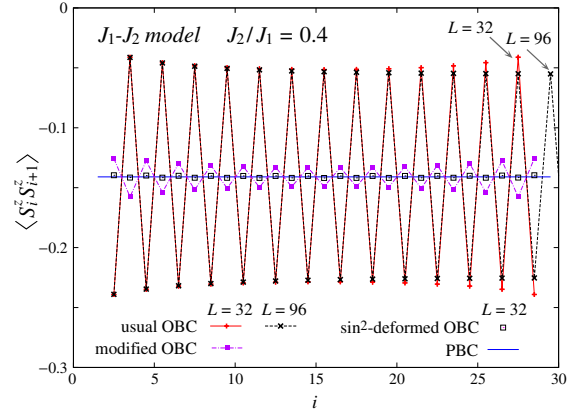}
\end{center}
\caption{ 
Comparison of local correlation ,$\langle S_i^z S_{i+1}^z \rangle$, in the 
$J_1$-$J_2$ model calculated under usual OBC, modified OBC, deformed OBC, 
and PBC, with $L=32$ and $m=200$. 
The convergence of the results regarding the $m$ dependence is confirmed. 
For modified OBC we use $J_{\rm edge}=0.475J$ on both ends. 
The results of OBC with $L=96$ is added to compare with the $L=32$ one.
}
\label{f2}
\end{figure} 
\begin{figure}[tbp]
\begin{center}
\includegraphics[width=7.5cm]{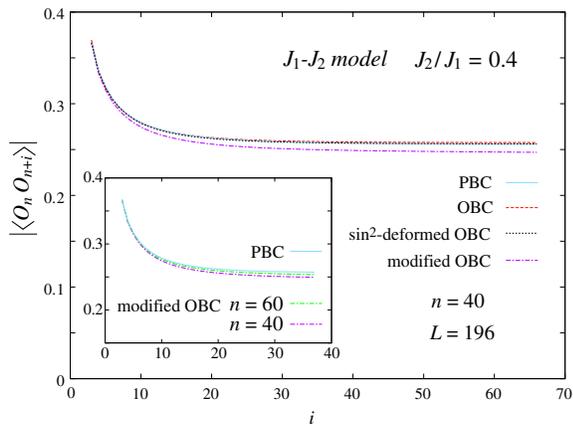}
\end{center}
\caption{
Comparison of two-point correlation functions, 
$|\langle O_n O_{n+i}\rangle|$, in the $J_1$-$J_2$ model with $L=196$ 
and $m=300$. 
Those of usual OBC, modified OBC, and deformed OBC start from $n=40$, 
and the PBC one from $n=1$. 
The convergence of the results regarding the $m$ dependence is confirmed. 
Inset shows the results of modified OBC with $n=60$ and 40. 
}
\label{f3}
\end{figure} 

\section{$J_1$-$J_2$ model}
\subsection{Dimer order}
Our first step is to examine how the order operators discussed in 
Sec.~\ref{sec2:tasaki} 
behave under the variation of boundary conditions 
when the symmetry-breaking long-range order is present in the bulk limit. 
For this purpose, we choose a $J_1$-$J_2$ model, whose Hamiltonian reads
\begin{equation}
{\mathcal H} = \sum_{j} \left( J_1\; {\bf S}_j \!\cdot \!{\bf S}_{j+1}+J_2\; 
{\bf S}_j \!\cdot \!{\bf S}_{j+2} 
\right),
\end{equation}
where $J_1>0$.
The exact ground state at $\alpha=J_2/J_1=0.5$, 
which is called the Majumdar-Ghosh (MG) point, 
is exactly represented by the product of local dimer states, 
and thus the dimer-dimer correlation is expected to show no decay for 
length scale large 
enough to neglect the boundary effect induced by OBC. 
The existence of dimer long-range order is established off the MG point in 
the region of 
$\alpha\gtrsim 0.24$.\cite{mg-point} 
\par
The nearest-neighbor spin-spin correlation operator is given as
\begin{equation}
D_i = {\bf S}_i \!\cdot \!{\bf S}_{i+1}, 
\end{equation}
where ${\bf S}_i$ is the spin operator of site $i$. 
We use the dimer operator
\begin{equation}
O_i = D_{i+1}-D_{i}= {\bf S}_{i+1}\!\cdot {\bf S}_{i+2} -{\bf S}_{i}\!\cdot {\bf S}_{i+1}
\label{def_o}
\end{equation}
to define the order parameter $\sigma$ in Eq.~(\ref{op2}). 
When the translational symmetry breaking dimer order is present, the 
two-point correlation function, $\langle O_i O_j \rangle$, 
remains finite in the bulk limit. 

\subsection{Boundary condition dependence}
\noindent
We now compare the results under three different types of boundary conditions 
at 
$J_2/J_1=0.4$ where the dimer long-range order is present in the bulk limit. 
Figure~\ref{f2} shows the site dependence of the local operator, 
$\langle S_i^z S_{i+1}^z\rangle=\langle D_i\rangle/3$. 
Under usual OBC, $\langle D_i\rangle$ shows twofold periodic oscillation 
with a large amplitude. 
The amplitude of oscillation is significantly suppressed, 
when the modified OBC is adopted with $J_{\rm edge}=0.475J$ on both edge bonds. 
Further, by the $\sin^2$ deformation, the oscillation is almost completely 
suppressed and the spatially uniform $\langle S_i^z S_{i+1}^z\rangle$ is 
obtained, 
which means that the translational symmetry is recovered. 
In fact, this deformation gives almost the identical results with that 
of the PBC 
given in the same figure. 
In all cases, the magnitude of $|\langle S_i^z S_{i+1}^z\rangle|$ has only 
small $i$ dependence. 
We also plot the results under OBC with $L=96$ to show that 
the $L$ dependence of the oscillation amplitude is significantly small, 
indicating that the distinct twofold oscillation remains after 
the finite-size scaling in OBC.
Therefore, there is no doubt that the extrapolated value of $\langle S_i^z 
S_{i+1}^z\rangle$ in the bulk limit, $\lim_{L\rightarrow\infty} \langle 
D_{L/2}\rangle$, depends severely on the boundary conditions. 
\par
Next, we show two point correlation functions, $|\langle O_n O_{n+i}\rangle|$, 
in Fig.~\ref{f3} under the same choices of boundary conditions used in 
Fig.~\ref{f2}. 
One finds that all cases almost asymptotically approach a constant value 
after $i\gtrsim 30$, in sharp contrast to the 
severely boundary-dependent $\langle D_i\rangle$. 
The slight difference for the modified boundary condition 
is due to a boundary effect of finite systems, and it
almost vanishes for large $n\gtrsim 60$ as shown in the
inset of Fig.~\ref{f3}. 
From this result, one can safely confirm that the extrapolated value of 
the two-point correlation function, 
$\lim_{i\rightarrow\infty} |\langle O_n O_{n+i} \rangle | $, 
is a reliable measure of symmetry-breaking long-range 
order for any choices of boundary conditions. 
%
\section{Kondo lattice model}
\subsection{Preliminary information}
The next example is devoted to the Kondo lattice model, which is one of 
the basic models for the heavy-fermionic systems\cite{klmexample}  
and is studied also as a prototype system to clarify the effects of coupling 
of localized spins and conduction electrons.\cite{klmexample2} 
The Hamiltonian in one dimension is given as
\[
{\mathcal H} = t\sum_{j} 
\left( c_{j\sigma}^\dagger c_{j+1\sigma}+ {\rm H.c.} \right)
+  J \sum_{j} {\bf S}_j \!\cdot \!{\bf s}_j, 
\]
where $c_{j\sigma}$ denotes the annihilation operator of the 
conduction electron at the $j$th site  
with spin $\sigma=\uparrow, \downarrow$, and
${\bf S}_j$ denotes the localized spin operator with $S=1/2$.
${\bf s}_j= \frac{1}{2}{\boldsymbol \tau}_{\sigma,\sigma'}
c_{j\sigma}^\dagger c_{j\sigma'}$
is the spin operator of the conduction electron with 
Pauli matrix ${\boldsymbol \tau}_{\sigma,\sigma'}$.
The conduction electrons hop with energy $t$ and
interact with localized spins through the Kondo exchange coupling $J$.
Its ground state in one dimension is considered as a 
Tomonaga-Luttinger liquid (TLL) in the weak-coupling region away from 
half filling.
In the strong-coupling region, each conduction electron forms a singlet 
with localized spins and behaves as a single hole, and the 
ferromagnetic metallic ground state is realized.\cite{shibata97} 
\par
A few years ago, however, there was a proposal by Xavier {\it et.al.} on the 
possible dimer phase in the region of $J/t \le 1.6$ at quarter filling,
\cite{xavier03} 
which was originally considered a paramagnetic TLL region. 
Based on the fact that $\langle O_{L/2}\rangle$ calculated under OBC with DMRG 
seems to remain finite after the size scaling, 
they claimed that the translational symmetry is broken in the bulk limit. 
We carried out similar DMRG study on the same model and reproduced their 
results, 
but with different conclusions based on the analyses of the  boundary 
effect.\cite{comment,comment2} 
In fact, the model at $0\le J/t \le 1$ is in a state which is extremely 
difficult to analyze even by DMRG established in one dimension.\cite{present} 

In the following we show in detail how the two-point correlations as well as 
the local quantities behave under the variation of boundary conditions, 
which has distinct differences from those of the dimer phase in 
the $J_1$-$J_2$ model. 

\begin{figure}[tbp]
\begin{center}
\includegraphics[width=7cm]{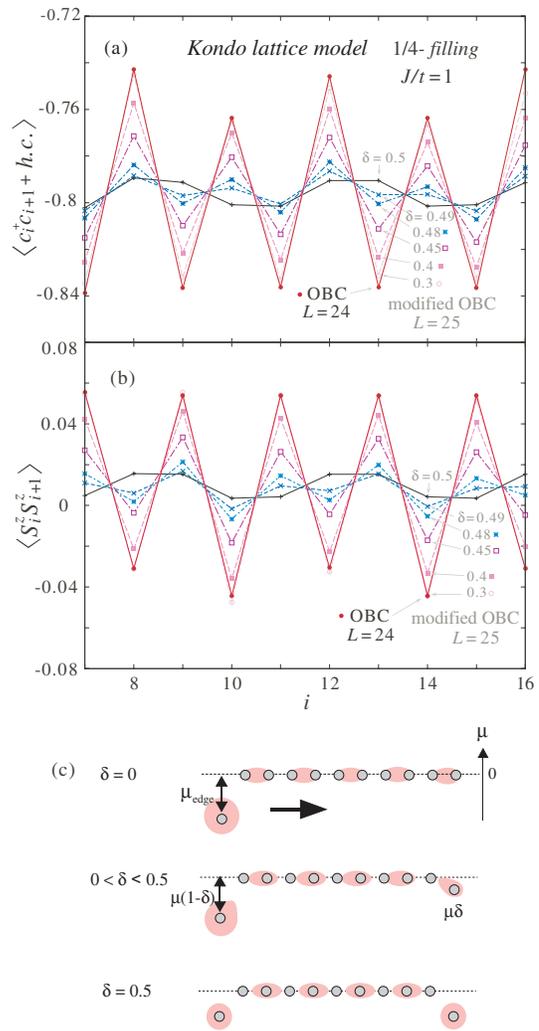}
\end{center}
\caption{
Spatial modulation of two local quantities, (a) bond kinetic energy, 
$\langle c_i^\dagger c_{i+1} + {\rm h.c.}\rangle$, 
and (b) nearest-neighbor correlation of localized spins, 
$\langle S_i^z S_{i+1}^z \rangle$, 
in the Kondo lattice model calculated with usual OBC with $L=24$ and 
modified OBC with $L=25$ with $m=600$. 
The convergence of the results regarding the $m$ dependence is confirmed. 
For modified OBC we use $\mu_{\rm edge}=\mu(1-\delta)$ and $\mu\delta$ on 
left and right edge site, respectively, with $\mu/t=-1.6$. 
(c) Schematic illustration of the electronic 
states under modified boundary condition 
by the chemical potential, $\mu_{\rm edge}$.
The size of the shaded circles represents the density of 
conduction electrons. 
}
\label{f4}
\end{figure} 
\begin{figure}[tbp]
\begin{center}
\includegraphics[width=7cm]{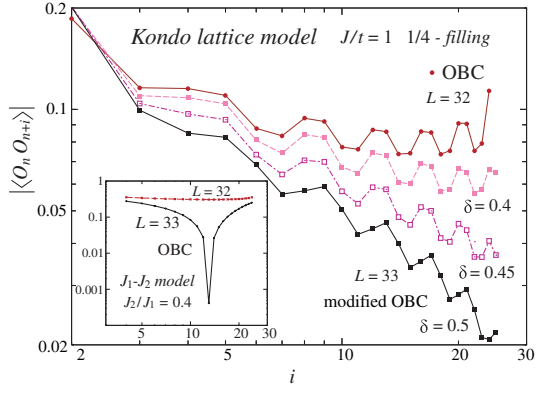}
\end{center}
\caption{
Two-point correlation functions of the dimer operator, 
$|\langle O_n O_{n+i}\rangle|$ (for fixed $n=5$) in the Kondo lattice model 
with $m=600$ (the $m$ dependence is found to be negligible for the present 
$L$'s). 
The usual OBC with $L=32$ and modified OBC with $L=33$ for several choices 
of $\delta$ ($\mu_{\rm edge}$ is the same as Fig.\ref{f4}) are compared. 
The inset shows the two correlation functions, $|\langle O_n O_{n+i}\rangle|$, 
of the $J_1$-$J_2$ model with usual OBC at $L=32$ and 33 with $n=5$, 
to be compared with those of the Kondo lattice model in the main panel. 
}
\label{f5}
\end{figure} 
\subsection{Modified open boundary}
In the KLM the open boundary condition can be modified by 
the chemical potential $\mu_{\rm edge}=\mu(1-\delta)$ and 
$\mu \delta$ on the left and the right edges, respectively. 
The value of $\mu/t$ is tuned to $-1.6$ in order to keep the 
density of charges at quarter filling within the accuracy of 
$10^{-3}$ away from the edge sites, and $\delta$ is 
introduced to shift the center of mass of electrons.
Figures~\ref{f4}(a) and \ref{f4}(b) show the site dependences of 
the bond kinetic energy $\langle c_i^\dagger c_{i+1} + {\rm h.c.}\rangle$ 
and the nearest-neighbor correlation of localized spins 
$\langle S_i^z S_{i+1}^z \rangle$ for several choices of $\delta$. 

Under the usual OBC with even $L$, both $\langle c_i^\dagger c_{i+1} + 
{\rm h.c.}\rangle$ 
and $\langle S_i^z S_{i+1}^z \rangle$ show large oscillation. 
The same oscillation is found when the modified boundary is adapted to 
the odd-$L$ system with $\delta=0$, 
namely when $\mu$ is placed on only one of the edge sites. 
From this result, one can understand that $\mu$ works to pin the extra 
charges off quarter filling, $N_e-(L-1)/2$, to the edge site, 
and the rest of the system remains quarter filled. 
Then, as $\delta$ is increased from 0 toward 0.5, the amplitude of 
oscillation is gradually suppressed. 
\par
Figure \ref{f4}(c) shows the schematic illustration of the effect of 
$\mu$ and $\delta$ on edge sites. 
The usual OBC with an even number of lattice sites pins the electron 
density to be the bond-centered spatial modulation along the 
one-dimensional chain.
When the boundary is modified as $0<\delta < 0.5$, the electrons on the
left and right edges have different hopping amplitude to their neighbors, 
which works to displace the spatial modulation of hoppings and to suppress the 
oscillation of hopping amplitude as we find in Fig.~\ref{f4}(a).
Since the hopping of electrons induces correlations between the local spins
through the Kondo exchange interaction, $\delta$ also modifies the local spin 
correlations as shown in Fig.~\ref{f4}(b).
This result arises the question of whether the oscillation of 
$\langle S_i^z S_{i+1}^z \rangle$ 
found in the OBC is an intrinsic property of the KLM.
To study the boundary condition dependence in detail,
we next analyze the two-point correlation function of the local spins. 
Figure \ref{f5} shows the dimer-dimer correlation of the local spins
$|\langle O_n O_{n+i}\rangle|$.
For usual OBC, $|\langle O_n O_{n+i}\rangle|$ saturates toward the finite 
value already at $i \sim 20$, which apparently suggests the existence 
of a dimer state. 
However, when the boundary is modified ($\delta >0$), the functional form 
changes significantly and starts to decay monotonously. 
\par
To examine whether the odd $L$ of the one-dimensional chain artificially 
suppresses the dimer correlation or not, 
we also calculate $|\langle O_n O_{n+i}\rangle|$ of the $J_1$-$J_2$ model 
under both even and 
odd $L$ starting from $n=5$ as in KLM, which is shown in the inset of 
Fig.\ref{f5}. 
In the case of the $J_1$-$J_2$ model with the true long-range dimer order
in the bulk limit, $|\langle O_n O_{n+i}\rangle|$ of the odd $L$ 
first decays and takes the minimum (or a dip) 
at around $i\sim 12$ which is near the system center, then shows an upturn, 
and finally approaches the value of the even $L$, which is almost an 
$i$-independent constant. 
The local decay (dip) of the dimer correlation is due to 
the kink structure formed by the phase shift 
$\pi$ between the two dimer orders starting from both ends: 
the misfit of odd $L$ and the twofold periodicity 
of the dimer order. 
Since the kink is located around the center of the system,
the dimer correlation gradually decreases toward the 
system center.
But for long distances beyond the center of the system, 
this kink is always located between the two dimer operators
$O_i$ and $O_j$, and the dimer correlation increases
up to its original value with an additional phase shift by $\pi$.

Such nonmonotonic correlation function is also expected for the
KLM of odd $L$, if the long range dimer order is present.
However, the results in Fig.\ref{f5} for finite $\delta$ show 
that the dimer correlation monotonically decreases for long distance
even for the case of symmetric edge potentials at $\delta=0.5$. 
This result suggests that the dimer order is not an essential 
property of the KLM at quarter filling at $J/t=1$.
The large $\delta$ dependence of electronic states [see Fig.~\ref{f4}(a)] and 
concurrently of the spatial structure of localized spins [Fig.~\ref{f4}(b)] 
are the sign that the electronic state is severely boundary condition 
dependent, and
one should analyze the finite-size scaling and its boundary condition 
dependence in order to determine the bulk property of the system safely.

\begin{figure}[tbp]
\begin{center}
\includegraphics[width=7cm]{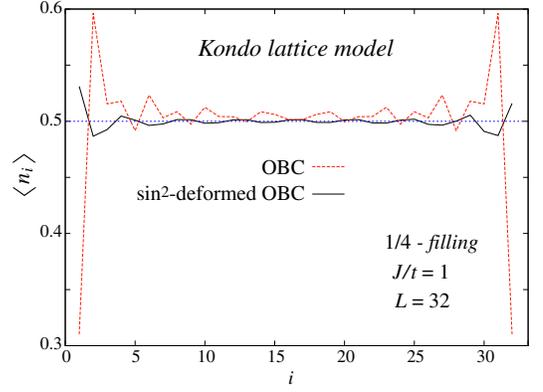}
\end{center}
\caption{
Site dependence of the electron density in the quarter-filled Kondo lattice 
model at $J/t=1$, under the usual OBC and $\sin^2$ deformed OBC 
with $L=32$ and $m=600$. 
}
\label{f6}
\end{figure} 
%
\begin{figure}[tbp]
\begin{center}
\includegraphics[width=7.5cm]{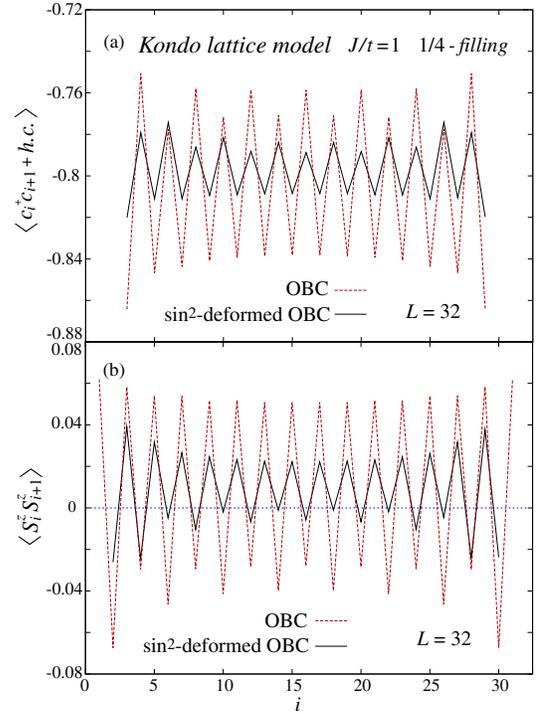}
\end{center}
\caption{
Site dependence of (a) bond kinetic energy, $\langle c_i^\dagger c_{i+1} 
+ {\rm h.c.}\rangle$, and (b) nearest-neighbor correlation of localized spins, 
$\langle S_i^z S_{i+1}^z \rangle$, in the quarter-filled Kondo lattice model 
at $J/t=1$, under the usual OBC and $\sin^2$ deformed OBC 
with $L=32$ and $m=600$. 
}
\label{f7}
\end{figure} 
%
\begin{figure}[tbp]
\begin{center}
\includegraphics[width=7cm]{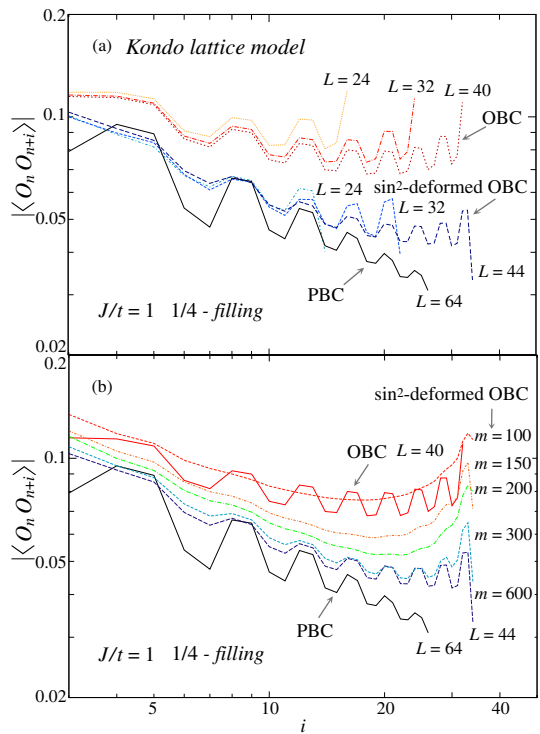}
\end{center}
\caption{
Two-point correlation functions of the dimer operator, $|\langle O_n 
O_{n+i}\rangle|$ (for fixed $n=5$) in the Kondo lattice model. 
(a) Series of results under usual OBC and deformed OBC by varying the 
system size $L$. 
(b) The variation of results of the deformed OBC ($L=44$) for several 
choices of $m$, to be compared with that of the usual OBC with $L=40$ 
and $m=600$. 
In both panels, the result of PBC with $L=64$ ($m=600$ for non-Abelian 
Hilbert space) is given for comparison. 
}
\label{f8}
\end{figure} 
\subsection{$\sin^2$ deformation and periodic boundary}
The final set of results is devoted to examining the 
effect of finite size $L$ and finite number of 
basis states $m$ in DMRG under the condition where 
the effect of the missing bond in OBC is as weakened as possible.
Based on the discussion in Sec.~\ref{sec2:deform}, we 
perform the calculation under the $\sin^2$ deformation,
which reproduces the translationally 
invariant wave function for the $J_1$-$J_2$ model.
\par
In the KLM, we add uniform chemical potential 
as $n$-independent $u(n)$ 
which is needed to keep the electron density at quarter filling, 
and deform the Hamiltonian according to Eq.~(\ref{deform}).
Figure~\ref{f6}(a) shows the comparison of the 
site dependence of electron density, 
$\langle n_i \rangle$, between usual OBC and deformed OBC. 
In usual OBC, the electron density significantly deviates 
from 0.5 near the system boundary, and as a result, the 
density around the center slightly exceeds quarter filling. 
Such deviation due to the boundary effect is suppressed in the 
deformed case, and the 
electron density becomes nearly site independent. 
Thus, the results of the latter are guaranteed to be 
less affected by the missing bond between the two edges.
\par
The comparison of the local quantities of two boundary 
conditions is given in Fig.~\ref{f7}. 
The oscillation amplitude of electron hopping, 
$\langle c^\dagger_i c_{i+1} + {\rm h.c.}\rangle$, 
and the nearest neighbor correlation of localized spins, 
$\langle S_i^z S_{i+1}^z\rangle$, are both suppressed
by the deformation of interaction. 
It is interesting to find that the mean value of 
$\langle S_i^z S_{i+1}^z\rangle$ 
remains a positive finite value even under the deformation. 
In usual OBC, $\langle S_i^z S_{i+1}^z\rangle$ has a large 
twofold oscillation between the comparable negative and 
positive values, and the local spin correlations are characterized 
by classical $q=\pi/2$ antiferromagnetic
state, $\uparrow\uparrow\downarrow\downarrow$. 
However, by the comparison of the two results, one can 
conclude that the antiferromagnetic correlation is only 
stabilized by the boundary effect (the missing bond in usual OBC).
The remaining ferromagnetic local correlation seems 
to be more intrinsic than the antiferromagnetic one in the $J/t=1$ KLM. 
In fact, the phase diagram given in Ref.\onlinecite{mcculloch02} 
indicates that system 
at this parameter is very close to the ferromagnetic 
phase transition. 
\par
Finally, we show in Figs. \ref{f8}(a) and \ref{f8}(b) the $L$ and 
$m$ dependencies of the two-point dimer correlation function, 
$|\langle O_n O_{n+i}\rangle|$.
The presented results show the following two important features:
The decay of $|\langle O_n O_{n+i}\rangle|$ is always more rapid for 
larger system size $L$ irrespective of whether we have usual 
OBC or deformed OBC. 
Having larger $m$ also leads to the rapid decay in the deformed OBC. 
Let us discuss the implication of these two features. 
In general, the accuracy of DMRG is guaranteed by 
asymptotically restricting the number of bases $m$ needed to 
reproduce the quantum states.
However, this technical advantage may sometimes bring in a 
by-product when $m$ is not taken large enough: an artificial 
suppression of intrinsic long-range quantum fluctuation which is 
required to suppress the boundary-induced symmetry braking.
If we focus on the value of $|\langle O_n O_{n+i} \rangle|$ at 
the particular length $i$ in Fig.~\ref{f8}(a), we always find 
smaller values for larger $L$. 
Since the calculation with larger $L$ includes the component of 
quantum fluctuation with larger length scale, 
this fact indicates that the dimer correlation should indeed 
be suppressed if the system includes a proper long-range quantum 
fluctuation present in the bulk limit. 
The same discussion holds for the $m$ dependence: by including a 
larger number of $m$, the more the fluctuation effect is included. 
If such fluctuation is intrinsic, the dimer correlation will be 
suppressed by the increase of $m$, which is in fact the case. 
\par
Let us briefly comment on the relation of the above $L$ and 
$m$ scaling with the entanglement entropy. 
In the systems with small $L$ and $m$, the number of eigenstates 
is limited to a small number. 
This smallness generally enhances the separation of discrete 
energy levels and suppresses long-range quantum fluctuations, 
which severely influences the accuracy of long-range two-point 
correlation functions near the critical point. 
Meanwhile, the long-range quantum fluctuation generates quantum 
entanglement between two regions in the system, 
so that the increase in $L$ and $m$ enhances the entanglement entropy. 
Therefore, not only the two-point correlation function but also 
the entanglement entropy shall serve as a measure of 
to what extent the long range quantum fluctuation is taken into 
account near the critical point. 
The detailed and quantitative analysis on the relation of 
entanglement entropy with the boundary effect 
is left as a future problem. 
\par
We finally note that the results of these $L$ and $m$ scaling 
asymptotically approach the one under the PBC in the same panel 
of Fig.~\ref{f8}. 
The $|\langle O_n O_{n+i}\rangle|$ of PBC clearly shows an 
algebraic decay characteristic of the Tomonaga-Luttinger liquid. 
As discussed in Ref. \onlinecite{nishino09} (see Sec.~II), 
the $\sin^2$ deformation
is one of the ways to recover the translational invariance, 
which is demonstrated in Sec.~III for the $J_1$-$J_2$ model. 
To recover perfectly the translational symmetry by the 
deformation in KLM at $J=1$ is rather out of scope in the 
present calculation,
because the on-site Kondo singlet correlation 
competes with the inter-site hopping, and the applicability
of the $\sin^2$ deformation is not clear.
However, even in such a difficult case, the analysis 
of the asymptotic behavior of $|\langle O_n O_{n+i}\rangle|$ 
gives us the strong indication that the true bulk property of 
the system approaches the one given under PBC, 
and that the dimer correlation is only induced 
by the boundary effect. 
The good coincidence of $|\langle O_n O_{n+i}\rangle|$ in 
Fig.~\ref{f5} and Fig.~\ref{f8} under different series of 
boundary conditions is regarded as its collateral 
evidence. 
\section{Summary and Discussion}
\noindent
To summarize, we made a case study analysis on the density matrix 
renormalization group calculation and proposed the systematic 
treatment to determine whether the symmetry breaking long range 
order exists or not. 
Since the two-point correlation function in the bulk limit
gives the lower bound of the symmetry-breaking 
long-range order parameter (see Sec.~\ref{sec2:tasaki}),  
we focused on the two-point correlation functions and 
systematically studied their boundary condition dependence. 
We demonstrated the calculations on the two contrasting cases, 
the $J_1$-$J_2$ model which has dimer long-range order, and the Kondo lattice 
model at quarter filling whose ground state has not reached a consensus 
yet due to numerical difficulty. 
\par
Four different types of boundaries are adopted and 
their effect on the expectation values of order operators are analyzed. 
In the usual OBC we always find oscillation of local quantities, e.g., 
the nearest neighbor spin-spin correlation 
$\langle S_i^z S_{i+1}^z \rangle$ or bond kinetic energy  
$\langle c_i^\dagger c_{i+1} + {\rm h.c.}\rangle$ 
oscillates with the largest amplitude. 
The modified OBC relaxes the oscillation to some extent under 
the variation of imbalance between potentials on both edges. 
Then, the deformation of interactions and PBC basically recover 
the homogeneity of local quantities. 
As the boundary effect fades out in such a way, the two-point 
correlation function also shows a systematic boundary
condition dependence, 
and thus the degree of inhomogeneity of the local quantities 
gives a measure of to what extent the boundary effect of OBC
is included in the wave function. 
Therefore, by examining the correspondence of local quantities 
and the two-point correlation function, one could analyze the overall 
features of 
the {\it boundary effect} as summarized in (III) of Sec.~\ref{sec2:tasaki}. 

Such boundary condition dependence is particularly distinct 
when the system is critical. 
When the system has symmetry-breaking long-range order (gapped), 
the two-point correlation function 
is insensitive to the boundary conditions. 
This fact is confirmed in the $J_1$-$J_2$ model, where the 
two-point correlation functions 
reach the same value after the finite-size scaling even when 
the corresponding local correlations take the significantly 
different value with each other. 
By sharp contrast, when the system lacks a distinct gap, 
the boundary condition severely affects the two-point correlation
function as we find in the Kondo lattice 
model:\cite{comment2} In OBC it seems to behave convex upward 
and asymptotically approach a finite value with increasing distance, 
which is easily transformed to the algebraic decay, once either 
of the modified OBC, $\sin^2$ deformation, or PBC is adopted. 
The boundary condition dependence of the two-point correlation function 
(Figs.\ref{f5} and \ref{f8}) is in 
good correspondence with that of the local quantities 
(Figs. \ref{f4} and \ref{f7}), 
and such boundary-dependent behavior is clearly different from that observed in 
the $J_1$-$J_2$ model. 
These results show that the dimer correlation observed in OBC is 
not an intrinsic property of the KLM and that {\it it is difficult to 
conclude that the symmetry-breaking long-range order does exist 
at quarter filling at $J/t=1$}. 
This fragile example which requires a difficult numerical 
treatment, provides us with a good lesson that numerical study, 
although it sometimes becomes as rigorous as analytical techniques, may 
have much analysis dependence, and one quite often needs care to 
understand the intrinsic physics lying behind. 
\par
We conclude that the DMRG calculation, which is often performed on OBC, 
sometimes reaches misleading conclusions. 
While OBC is very useful to analyze the detailed nature 
of the well-known ground state, 
we argue that to have a reliable conclusion on the determination of the 
ground state itself, the examination of the boundary condition dependence 
on the two-point correlation should be performed.
\par
This work is supported by Grant-in-Aid for Scientific Research (No.~21110522, 
No.~19740218, and No.~20102008) from MEXT, Japan.



\begin{references}
\bibitem{kink} In Ref.\onlinecite{xavier08}, 
the authors claim that $\langle {\bf S}_i\cdot {\bf S}_{i+1} \rangle$ near the kink structure in the odd-$L$ is suppressed, so that the odd $L$ system 
we adopt for modified OBC is improper. We comment that the proper analysis given in the paper overcomes this issue.
\bibitem{present}
At present, in KLM at quarter filling, 
we confirmed that the charge gap closes at $J/t \gtrsim 0.8$, 
and there must exist a critical phase transition point at $J/t<0.8$.
\end{references}
\end{document}